\documentclass[12pt]{article}
\usepackage{amssymb,amsmath,epsfig,graphicx}
\graphicspath{ {images/} }

\begin{document}
\title{\bf Some Exact Solutions in $f(\mathcal{G},T)$ Gravity via Noether Symmetries}

\author{M. Farasat Shamir $^{(1)}$\thanks{farasat.shamir@nu.edu.pk},
and Mushtaq Ahmad $^{(2)}$ \thanks{mushtaq.sial@nu.edu.pk}.\\\\ $^{(1)}$National University of Computer and
Emerging Sciences,\\ Lahore Campus, Pakistan, \\$^{(2)}$National University of Computer and
Emerging Sciences,\\ Chiniot-Faisalabad Campus, Pakistan.}

\date{}

\maketitle
\begin{abstract}

This paper is devoted to investigate the recently proposed modified Gauss-Bonnet $f(\mathcal{G},T)$ gravity, with $\mathcal{G}$, the Gauss-Bonnet term, coupled with ${T}$, the trace of energy-momentum tensor. We have used the Noether symmetry methodology to discuss some cosmologically important $f(\mathcal{G},T)$ gravity models with anisotropic background. In particular, the Noether symmetry equations for modified  $f(\mathcal{G},T)$ gravity are reported for locally rotationally symmetric Bianchi type $I$ universe. Explicitly, two models have been proposed to explore the exact solutions and the conserved quantities. It is concluded that the specific models of modified Gauss-Bonnet gravity may be used to reconstruct $\Lambda$CDM cosmology without involving any cosmological constant.

\end{abstract}


\section{Introduction}

Recent observations through huge telescopes and satellites have confirmed the evidence for an evolving universe which is expanding. The interpretations from the supernova experiments, the data from Wilkinson Microwave Anisotropy Probe (WMAP) and Sloan Digital Sky Surveys (SDSS) lead us towards the remarkable conclusion that this expansion is in accelerating mode \cite{1R}-\cite{2R}. Cosmological models are facing serious issues which can be summed up as the problems of the dark matter and dark energy. It is believed that so-called factor of dark matter and dark energy which is the 70\% of the total energy-mass of the universe is causing this acceleration. There is another viewpoint that the modifications of General Relativity (GR) are behind this accelerating expansion of this universe. Investigating all the possible reasons, including the dark energy problem, for the accelerating universe expansion would be one of the most focused areas of research followed by a series of experiments and several surveys, for the many years to come. No doubt, GR has been a great success in the last century but it could not properly address the problems of dark matter, dark energy, initial singularity, late-time cosmic acceleration, and the flatness issues. As an alternative to GR, different modified theories of gravity have been presented by the researchers to unveil these unsolved problems, which are believed to be the real cause of this accelerating expansion of the universe. Inspired by the original theory, a variety of modified theories of gravity like $f(R),~f(R,T),~f(\mathcal{G}),~f(R,\mathcal{G})$ and $f(\mathcal{G},T)$ have been structured \cite{cylndr}-\cite{sharif.ayesha}.

In particular, modified theories of Gauss-Bonnet (GB) gravity have attracted much attention of the researchers in recent years \cite{17}-\cite{18}. Late-time cosmic acceleration may be caused due to the existence of de-Sitter point of $f(\mathcal{G})$ gravity, $\mathcal{G}$ be the GB invariant defined as
\begin{equation}
\mathcal{G}=R^{2}-4R_{\gamma\delta}R^{\gamma\delta}+R_{\gamma\delta\alpha\beta}R^{\gamma\delta\alpha\beta},
\end{equation}
where $R_{\gamma\delta\alpha\beta}$ is the Reimann tensor, $R_{\gamma\delta}$ is the Ricci tensor and $R$ is the Ricci scalar. The notable attribute of this theory is that the participation of GB term may prevent uncertain contributions and uniforms the gravitational action \cite{Chiba}. The theory has been modified further by introducing scalar curvature along with GB invariant which is named as $f(R,\mathcal{G})$ gravity \cite{NJO6}. It is shown that $\Lambda$CDM epoch can be reconstructed using modified GB theories of gravity \cite{CDM}. A descent amount of work has been published so far in these theories \cite{19}-\cite{Sebastiani}. In a recent paper \cite{sharif.ayesha}, Sharif and Ikram introduced a new modified theory known as $f(\mathcal{G},T)$ gravity that includes the trace of the energy-momentum tensor in the function. They also determined that the massive test particles follows non-geodesic lines of geometry due to the presence of extra force and examined the energy conditions for Friedmann Robertson Walker (FRW) universe.
The same authors \cite{sharif.ayesha1} reproduced the cosmic evolution corresponding to de Sitter universe, power-law solutions and phantom/non-phantom eras in this theory using reconstruction techniques. For some particular choices of $f(\mathcal{G},T)$ gravity models, it is anticipated that this theory may explain the late-time cosmic acceleration.

The exact solutions of the differential equations have been calculated with the help of the symmetry methods of approximations. The complexity involved in a system of non-linear equations is minimized through these by finding the undetermined variables of equations. The Noether symmetries act like gadgets that provide the solutions and in addition to this, their existence gives suitable conditions so that one can choose physical models of the universe compatible with cosmological observations. Many authors have used Noether symmetries to investigate the cosmology in different contexts \cite{Jamil23}-\cite{noetherfG2}.
Using approximate symmetries, Sharif and Waheed \cite{Sharif13} re-scaled the energy of stringy charged black hole solutions. Using Noether symmetries, Kucukakca \cite{Kucuka16} found the exact solutions of Bianchi type-$I$ model. Jamil et al. \cite{Jamil17} discusses $f(\mathcal{T})$, where $\mathcal{T}$ is the torsion scalar, specifically for the phantom and quintessence models using the Noether symmetry approach. Sharif and Shafique \cite{sharif19} examined Noether symmetries in a modified scalar-tensor gravity. The exact solutions in $f(R)$ gravity were also studied using Noether symmetries methods for FRW spacetime \cite{Cap21}. In a recent paper \cite{shamir.ahmad}, we investigated $f(\mathcal{G},T)$ gravity using Noether symmetry approach. Two specific models were studied to determine the conserved quantities and it was concluded that the
well known deSitter solution could be reconstructed for some specific choice of $f(\mathcal{G},T)$ gravity model. Thus, it seems interesting to explore further the modified $f(\mathcal{G},T)$ gravity.

In this paper, our main focus is to investigate $f(\mathcal{G},T)$ gravity with anisotropic background. We have considered the locally rotationally
symmetric (LRS) Bianchi type $I$ spacetime for this purpose. Moreover, we adopt Noether symmetry approach for the present analysis due to the complicated and highly non-linear nature of the field equations. The organization of this paper is as follows: In section \textbf{2}, we provide the basic framework for $f( \mathcal{G},T)$ gravity. Section \textbf{3} presents the Noether equations of LRS Bianchi type $I$ universe model for $f(\mathcal{G},T)$ gravity. Reconstruction of some important cosmological solutions and graphical analysis is given in section \textbf{4}. The last section is comprised of a brief outlook of the paper.

\section{ Preliminary Formalism of $f(\mathcal{G},T)$ Gravity }

The general action for the modified $f(\mathcal{G},T)$ is \cite{sharif.ayesha},
 \begin{equation}\label{action}
\mathcal{A}= \frac{1}{2{\kappa}^{2}}\int d^{4}x
\sqrt{-g}[R+f(\mathcal{G},\mathrm{\textit{T}})]+\int
d^{4}x\sqrt{-g}\mathcal{L}_{M},
\end{equation}
where the function $f( \mathcal{G},T)$ is comprised of the GB term $\mathcal{G}$ and the trace of the energy-momentum tensor $T$, $\kappa$ is the coupling constant, $g$ is the determinant of the metric tensor, $R$ is the Ricci Scalar, and $\mathcal{L}_{M}$ denotes the matter part of the Lagrangian. The energy-momentum tensor denoted by $\mathrm{\textit{T}}_{\zeta\eta}$ can be given as
\begin{equation}\label{emt}
\mathrm{\textit{T}}_{\zeta\eta}=-\frac{2}{\sqrt{-g}}\frac{\delta(\sqrt{-g}\mathcal{L}_{M})}{\delta
g^{\zeta\eta}}.
\end{equation}
However, the metric dependent energy-momentum tensor will have the form
\begin{equation}\label{emt1}
\mathrm{\textit{T}}_{\zeta\eta}=g_{\zeta\eta}\mathcal{L}_{M}-2\frac{\partial\mathcal{L}_{M}}{\partial
g^{\zeta\eta}}.
\end{equation}
The following field equations are obtained by varying Eq.(\ref{action}) with respect to the metric tensor.
\begin{eqnarray}\nonumber
G_{\zeta\eta}&=&\kappa^{2}\mathrm{\textit{T}}_{\zeta\eta}-[2Rg_{\zeta\eta}\nabla^{2}-
2R\nabla_{\zeta}\nabla_{\eta}-4g_{\zeta\eta}R^{\mu\nu}\nabla_{\mu}\nabla_{\nu}-
4R_{\zeta\eta}\nabla^{2}+\\\nonumber
&&4R^{\mu}_{\zeta}\nabla_{\eta}\nabla_{\mu}+4R^{\mu}_{\eta}\nabla_{\zeta}\nabla_{\mu}+4R_{\zeta\mu\eta\nu}\nabla^{\mu}\nabla^{\nu}]f_{\mathcal{G}}+
\frac{1}{2}g_{\zeta\eta}f-[\mathrm{\textit{T}}_{\zeta\eta}+\Theta_{\zeta\eta}]\times\\
&&f_{\mathrm{\textit{T}}}-[2RR_{\zeta\eta}-4R^{\mu}_{\zeta}R_{\mu\eta}-4R_{\zeta\mu\eta\nu}R^{\mu\nu}+2R^{\mu\nu\delta}_{\zeta}R_{\eta\mu\nu\delta}]
f_{\mathcal{G}},\label{4_eqn}
\end{eqnarray}
where $\Box=\nabla^{2}=\nabla_{\zeta}\nabla^{\zeta}$ stands for the
d'Alembertian operator, ${G}_{\zeta\eta}=R_{\zeta\eta}-\frac{1}{2}g_{\zeta\eta}R$ is the
Einstein tensor, $\Theta_{\zeta\eta}= g^{\mu\nu}\frac{\delta
\mathrm{\textit{T}}_{\mu\nu}}{\delta g_{\zeta\eta}}$, $f\equiv f(\mathcal{G},T)$, $f_{\mathcal{G}}\equiv\frac{\partial f (\mathcal{G},\mathrm{\textit{T}})}{\partial
\mathcal{G}}$, and $f_{\mathrm{\textit{T}}}\equiv\frac{\partial
f(\mathcal{G},\mathrm{\textit{T}})}{\partial \mathrm{\textit{T}}}$.
Einstein equations are reproduced by putting
$f(\mathcal{G},\mathrm{\textit{T}})=0$ whereas field equations for
$f(\mathcal{G})$  are obtained simply by replacing
$f(\mathcal{G},\mathrm{\textit{T}})$ with $f(\mathcal{G})$ in Eq.($\ref{4_eqn}$).
Given below is the trace of Eq.($\ref{4_eqn}$) which gives us an important equation as it can be used to find the corresponding cosmological $f(\mathcal{G},\mathrm{\textit{T}})$ models
\begin{equation}\label{5_eqn}
R+\kappa^{2}\mathrm{\textit{T}}-(\mathrm{\textit{T}}+\Theta)f_{\mathrm{\textit{T}}}+2f+2\mathcal{G}f_{\mathcal{G}}
-2R\nabla^{2}f_{\mathcal{G}}+4R^{\zeta\eta}\nabla_{\zeta}\nabla_{\eta}f_{\mathcal{G}}=0.
\end{equation}
The covariant divergence of Eq.(\ref{4_eqn}) is given as
\begin{equation}\label{div}
\nabla^{\zeta}T_{\zeta\eta}=\frac{f_{\mathrm{\textit{T}}}}
{\kappa^{2}-f_{\mathrm{\textit{T}}}}\bigg[(\mathrm{\textit{T}}_{\zeta\eta}+\Theta_{\zeta\eta})
\nabla^{\zeta}(\text{ln}f_{\mathrm{\textit{T}}})+
\nabla^{\zeta}\Theta_{\zeta\eta}-
\frac{g_{\zeta\eta}}{2}\nabla^{\zeta}T\bigg].
\end{equation}
Eq.(\ref{div}) indicates that the conservation equation of energy momentum tensor is not verified in usual Einstein's theory.
However, the standard conservation equation for energy momentum tensor can be established by putting some constraints to Eq.(\ref{div}).
In this paper, we restrict ourselves to LRS Bianchi I space-time:
\begin{equation}\label{13_eqn}
ds^{2}=d{t}^{2}-A^{2}(t)dx^{2}-B^{2}(t)[d{y}^{2}+d{z}^{2}],
\end{equation}where $A$ and $B$ are the cosmic scale factors.
Corresponding  expressions for the Ricci scalar and GB term are given respectively as
\begin{equation}\label{Ricci2}
R=-2\bigg(\frac{\ddot{A}}{A}+2\frac{\ddot{B}}{B}+2\frac{\dot{A}\dot{B}}{AB}+\frac{\dot{B^{2}}}{B^{2}}\bigg),~~~
\mathcal{G}=8\bigg(\frac{\ddot{A}\dot{B}^{2}}{AB^{2}}+2\frac{\dot{A}{\dot{B}\ddot{B}}}{AB^{2}}\bigg),
\end{equation}
where the dot gives the derivative with respect to time $t$. The standard matter energy-momentum tensor is defined as
\begin{equation}\label{6}
T_{\alpha\beta}=(\rho + p)u_\alpha u_\beta-pg_{\alpha\beta},
\end{equation}
satisfying the equation of state (EoS) parameter
\begin{equation}\label{eos}
\omega=\frac{p}{\rho},
\end{equation}
where $\rho$ and $p$ denote energy density and
pressure of the fluid respectively.
It is mentioned here that the EoS parameter $\omega$ is important as it may describe the different epochs of the accelerating universe. For $-1<\omega\leq-1/3$, the dark energy (DE) phase is divided by quintessence epoch whereas $\omega<-1$ and $\omega=-1$ corresponds to phantom and cosmological constant $\Lambda$ eras respectively. It is very difficult to determine the exact solutions of extremely non-linear partial differential equations (PDE's) as the modified field equations for LRS Bianchi type $I$ spacetime are of fourth order with unknowns including  $f(\mathcal{G}, \mathrm{\textit{T}})$.
The benefit of exact solutions in modified gravity has significant importance, especially in the study of phase transitions and recent phenomenon of accelerated expansion of universe. Noether symmetries can be used to find the viable cosmological models and as a result, some solutions with physical importance can be established. 

\section{Noether Symmetries and $f(\mathcal{G},T)$ Gravity}

Noether symmetries have an important role in cosmology as they help to find the solutions of system of non-linear field equations. The uniqueness of the vector field in the tangent space can be established through the existence of Noether symmetry approach. More conserved quantities can be recovered using this approach.
For the case of perfect fluid, we rewrite the action
(\ref{action}) as
\begin{equation}\label{32_eqn}
\mathcal{A}=\int dt\sqrt{-g}[R+f(\mathcal{G},\mathrm{\textit{T}})-\mu_{1}(\mathcal{G}-\mathcal{\bar{G}})-\mu_{2}(\mathrm{\textit{T}}-\bar{T})+\mathcal{L}_{M}].
\end{equation}
Here $\bar{\mathrm{\textit{T}}}$ and $\bar{\mathcal{G}}$ stand for dynamical
constraints, while the Lagrange multipliers $\mu_{1}$ and $\mu_{2}$ are calculated as
\begin{equation*}\label{33_eqn}
\begin{split}
\mu_{1}=f_{\mathcal{G}}(\mathcal{G},\mathrm{\textit{T}}),~~
\mu_{2}=f_{\mathrm{\textit{T}}}(\mathcal{G},\mathrm{\textit{T}}).
\end{split}
\end{equation*}
Since matter Lagrangian may have different definitions so here,
we consider $\mathcal{L}_{M}=-p(B)$. Moreover, we assume a constant ratio of shear and expanding scale factors, which gives   $A=B^m$, where $m$  is an arbitrary real number and for the sake of non-trivial solutions, we consider $m\neq0,1$ \cite{ratio} . Hence, after integration by parts, the point-like Lagrangian becomes
\begin{equation}
\begin{split}
\mathcal{L}(B,R,\mathcal{G},\mathrm{\textit{T}},\dot{B},\dot{R},\dot{\mathcal{G}},\dot{\mathrm{\textit{T}}})=
B^{m+2}[R+f-\mathcal{G}f_{\mathcal{G}}-f_{\mathrm{\textit{T}}}\{\mathrm{\textit{T}}-\rho(B)-3p(B)\}-
\\p(B)]-8mB^{m-1}\dot{B}^{3}\dot{\mathcal{G}}f_{\mathcal{GG}}-
8mB^{m-1}\dot{B}^{3}\dot{\mathrm{T}}f_{\mathcal{G}\mathrm{T}}.\\\label{34_eqn}
\end{split}
\end{equation}
We consider the vector field and it's first prolongation  respectively as \cite{Olver}
\begin{equation}\label{204_eqn}
\mathrm{\textit{W}}=\zeta(t,u^{j})\frac{\partial}{\partial
t}+\xi^{i}(t,u^{j})\frac{\partial}{\partial u^{j}},
\end{equation}
\begin{equation}\label{24_eqn}
\mathrm{\textit{W}}^{[1]}=\mathrm{\textit{W}}+(\xi^{i}_{~,t}+\xi^{i}_{~,j}~\dot{u}^{j}-\zeta_{~,t}~\dot{u}^{i}-\xi_{~,j}~
\dot{u}^{j}\dot{u}^{i})\frac{\partial}{\partial\dot{u}^{j}},
\end{equation}
where $\zeta$ and $\xi$ are the coefficients of the generator, $u^i$ provides the $n$ number of positions.
The Noether gauge symmetry given by the vector field $\textit{W}$
\begin{equation}\label{25_eqn}
\mathrm{\textit{W}}^{[1]}\mathcal{L}+(D\zeta)\mathcal{L}=DG(t,u^{i})
\end{equation}
is preserved and the gauge term is denoted by $G(t,u^i)$ and operator $D$ is defined as
\begin{equation*}\label{26_eqn}
D=\frac{\partial}{\partial
t}+\dot{u}^{i}\frac{\partial}{\partial u^{i}}.
\end{equation*}
The Euler-Lagrange equations are given by
\begin{equation}\label{29_eqn}
\frac{\partial \mathcal{L}}{\partial
u^{i}}-\frac{d}{dt}\Bigg(\frac{\partial \mathcal{L}}{\partial
\dot{u}^{i}}\Bigg)=0.
\end{equation}
Contraction of Eq.(\ref{29_eqn}) with some unknown function $\phi^{i}\equiv\phi^{i}(u^{j})$ yields
\begin{equation}
\phi^{i}\Big(\frac{\partial \mathcal{L}}{\partial
u^{i}}-\frac{d}{dt}\Big(\frac{\partial \mathcal{L}}{\partial
\dot{u}^{i}}\Big)\Big)=0.\label{12}
\end{equation}
It can be easily verified that
\begin{equation}
\frac{d}{dt}\Big(\phi^{i}\frac{\partial \mathcal{L}}{\partial
\dot{u}^{i}}\Big)-\Big(\frac{d}{dt}\phi^{i}\Big)\frac{\partial
\mathcal{L}}{\partial \dot{u}^{i}}
=\phi^{i}\frac{d}{dt}\Big(\frac{\partial \mathcal{L}}{\partial
\dot{u}^{i}}\Big).\label{13}
\end{equation}
Use of the Eq.(\ref{13}) in Eq.(\ref{12}) results
\begin{equation}\label{abc}
L_\textit{W}\mathcal{L}
=\phi^{i}\frac{\partial \mathcal{L}}{\partial
u^{i}}+\Big(\frac{d}{dt}\phi^{i}\Big)\frac{\partial
\mathcal{L}}{\partial
\dot{u}^{i}}=\frac{d}{dt}\Big(\phi^{i}\frac{\partial
\mathcal{L}}{\partial \dot{u}^{i}}\Big),
\end{equation}
where $L$ represents the Lie derivative along the vector field. The
existence of the Noether symmetries would be possible only if the
Lie derivative of the Lagrangian becomes zero, i.e., the condition
\begin{equation*}\label{28_eqn}
L_{\mathrm{\textit{W}}}\mathcal{L}=0.
\end{equation*}
Since the Lagrangian $\mathcal{L}$ remains unchanged along the
vector field $\textit{W}$, as a result, the definition of Noether
current turns out to be \cite{Scap2 Note}
\begin{equation}
j^{t}=\Big(\phi^{i}\frac{\partial \mathcal{L}}{\partial
\dot{u}^{i}}\Big),\label{16}
\end{equation}
and in order to conserve the Noether current, we must have
\begin{equation}\label{17}
j^{t}_{,t}=0.
\end{equation}
The Euler-Lagrangian equations (\ref{29_eqn}) in this case are found as
\begin{eqnarray}\nonumber
&&(m+2)B^{m+1}\bigg[R+f-\mathcal{G}f_{\mathcal{G}}-f_{T}\{T-(\rho(B)-3p(B))\}-p(B)\bigg]+\\\nonumber
&&B^{m+2}\bigg[f_{T}(\rho_{,B}(B)-3p_{,B}(B))-p_{,B}(B)\bigg]+24m\bigg[f_{\mathcal{GG}}\Big((m-1)B^{m-2}\dot{B}^{3}\dot{\mathcal{G}}+\\\nonumber
&&2B^{m-1}\dot{B}\ddot{B}\dot{\mathcal{G}}+B^{m-1}\dot{B}^{2}\ddot{\mathcal{G}}\Big)
+f_{\mathcal{G}T}\Big((m-1)B^{m-2}\dot{B}^{3}\dot{T}+2B^{m-1}\dot{B}\ddot{B}\dot{T}+\\\label{35_eqn}
&&B^{m-1}\dot{B}^{2}\dot{T}\Big)+2B^{m-1}\dot{B}^{2}\dot{\mathcal{G}}\dot{T}f_{\mathcal{GG}T}
+B^{m-1}\dot{B}^{2}\Big(\dot{\mathcal{G}}^{2}f_{\mathcal{GGG}}+\dot{T}^{2}f_{\mathcal{G}TT}\Big)\bigg]=0,
\end{eqnarray}
\begin{equation}\label{36_eqn}
B^{m+2}\bigg[-\mathcal{G}f_{\mathcal{GG}}-Tf_{\mathcal{G}T}+f_{\mathcal{G}T}\big(\rho(B)-3p(B)\big)\bigg]
+8mf_{\mathcal{GG}}\Big(3B_{m-1}\dot{B}^{2}\ddot{B}+(m-1)B^{m-2}\dot{B}^{4}\Big)=0,
\end{equation}
\begin{equation}\label{37_eqn}
B^{m+2}\bigg[-\mathcal{G}f_{\mathcal{G}T}-Tf_{TT}+f_{TT}\big(\rho(B)-3p(B)\big)\bigg]
+8mf_{\mathcal{G}T}\Big(3B_{m-1}\dot{B}^{2}\ddot{B}+(m-1)B^{m-2}\dot{B}^{4}\Big)=0.
\end{equation}
Also the corresponding vector field using Eq.(\ref{24_eqn}) takes the form
\begin{equation}\label{37_eqn}
\mathrm{\textit{W}}=\alpha\frac{\partial}{\partial B}+\beta\frac{\partial}{\partial
R}+\gamma\frac{\partial}{\partial
\mathcal{G}}+\delta\frac{\partial}{\partial
\mathrm{\textit{T}}}+\dot{\alpha}\frac{\partial}{\partial
\dot{B}}+\dot{\beta}\frac{\partial}{\partial
\dot{R}}+\dot{\gamma}\frac{\partial}{\partial
\dot{\mathcal{G}}}+\dot{\delta}\frac{\partial}{\partial
\dot{\mathrm{\textit{T}}}},
\end{equation}
where $\alpha,~\beta,~\gamma$ and $\delta$ are functions of $B,~R,~\mathcal{G}$ and $\mathrm{\textit{T}}$.
Now using Lagrangian $(\ref{34_eqn})$ and Noether equation $(\ref{25_eqn})$ without the gauge term, an over-determined system of PDE's is obtained:
\begin{equation}\label{14_eqn}
\begin{split}
&(m-1)\alpha B^{m-2}f_{\mathcal{GG}}+ \gamma
B^{m-1}f_{\mathcal{GGG}}+\delta
B^{m-1}f_{\mathcal{GG}T}\\&+3B^{m-1}f_{\mathcal{GG}}\frac{\partial
\alpha}{\partial B}+B^{m-1}\frac{\partial \gamma}{\partial
\mathcal{G}}f_{\mathcal{GG}}+B^{m-1}\frac{\partial
\delta}{\partial \mathcal{G}}f_{\mathcal{G}T}=0,
\end{split}
\end{equation}
\begin{equation}\label{15_eqn}
\begin{split}
&(m-1)\alpha B^{m-2}f_{\mathcal{G}T}+ \gamma
B^{m-1}f_{\mathcal{GG}T}+\delta
B^{m-1}f_{\mathcal{G}TT}\\&+B^{m-1}f_{\mathcal{G}T}\frac{\partial
\alpha}{\partial B}+B^{m-1}\frac{\partial \gamma}{\partial
T}f_{\mathcal{GG}}+B^{m-1}\frac{\partial \delta}{\partial
T}f_{\mathcal{G}T}=0,
\end{split}
\end{equation}
\begin{equation}\label{16_eqn}
\frac{\partial \gamma}{\partial
R}f_{\mathcal{GG}}+\frac{\partial \delta}{\partial
R}f_{\mathcal{G}T}=0,~~~\frac{\partial \gamma}{\partial
B}f_{\mathcal{GG}}+\frac{\partial \delta}{\partial
B}f_{\mathcal{G}T}=0,
\end{equation}
\begin{equation}\label{20_eqn}
\frac{\partial \alpha}{\partial
\mathcal{G}}f_{\mathcal{G}T}+\frac{\partial
\alpha}{\partial T}f_{\mathcal{GG}}=0,
\end{equation}
\begin{equation}\label{21_eqn}
\frac{\partial \alpha}{\partial
\mathcal{G}}f_{\mathcal{GG}}=0,~~~\frac{\partial \alpha}{\partial R}f_{\mathcal{GG}}=0,~~~
\frac{\partial \alpha}{\partial T}f_{\mathcal{G}T}=0,~~~\frac{\partial \alpha}{\partial R}f_{\mathcal{G}T}=0,
\end{equation}
\begin{equation}\label{23_eqn}
\begin{split}
&\alpha(m+2)B^{m+1}\big[R+f(\mathcal{G},T)-\mathcal{G}f_{\mathcal{G}}-f_{T}(T-\{\rho(B)-3p(B)\})-p(B)\big]\\&+B^{m+2}\beta
+\alpha
B^{m+2}\big[-f_{T}\big(\rho_{,B}(B)-3p_{,B}\big)-p(B)_{,B})\big]+\gamma
B^{m+2}[-\mathcal{G}f_{\mathcal{GG}}-\\&f_{\mathcal{G}T}(T-\{\rho(B)-3p(B)\})]+\delta
B^{m+2}[-Tf_{TT}+f_{TT}(\rho(B)-3p(B))]=0.
\end{split}
\end{equation}
Using Eq.(\ref{17}), conservation equation for Noether charge
becomes
\begin{eqnarray}\label{18}
\begin{split}
&\frac{d}{dt}\Bigg[\alpha\Big\{\frac{\partial}{\partial
\dot{B}}\big[8mB^{m-1}\dot{B}^{3}\big(\dot{\mathcal{G}}f_{\mathcal{GG}}+
\dot{\mathrm{T}}f_{\mathcal{G}\mathrm{T}}\big)\big]\Big\}\\&+
\gamma\frac{\partial}{\partial
\dot{\mathcal{G}}}(8mB^{m-1}\dot{B}^{3}\dot{\mathcal{G}}f_{\mathcal{GG}})+\delta\frac{\partial}{\partial
\dot{\mathrm{T}}}(8mB^{m-1}\dot{B}^{3}\dot{\mathrm{T}}f_{\mathcal{G}\mathrm{T}})\Bigg]=0.
\end{split}
\end{eqnarray}
Since the system of PDE's (\ref{14_eqn}-\ref{23_eqn}) is complicated and highly non-linear, so we make different assumptions for $f(\mathcal{G},T)$ gravity models to investigate a solution. We first assume that $f_{\mathcal{G}\mathcal{G}}=0$. Using the
Eqs.(\ref{14_eqn}-\ref{23_eqn}), a trivial solution is obtained $\alpha=0$, $\beta=0$, $\gamma=0$, and $\delta=0$.
Additionally, the conservation equation (\ref{18}) is also
satisfied in this case. For a non-trivial solution, we have to
consider $f_{\mathcal{G}\mathcal{G}}\neq0$. Therefore, as a next step, we consider
$f_{\mathcal{G}\mathcal{\textit{T}}}=0$ and
$f_{\mathcal{G}\mathcal{G}}\neq0$. This choice provides us with $f(\mathcal{G},\textit{T})=a_{0}\mathcal{G}^{2}+b_{0}\textit{T}^{2}$, where $a_{0}$ and
$b_{0}$ are the arbitrary constants.
Here, using the Noether equations we get $\alpha=0=\gamma$,
$\delta=c_{1}$ and $\beta=c_{1}\textit{T}+c_{2}$. Thus the symmetry
generator takes the form
\begin{equation}
\textit{W}=(c_{1}\textit{T}+c_{2})\frac{\partial}{\partial R}+c_{1}\frac{\partial}{\partial\textit{T}}.
\end{equation}
In this case, the conservation equation (\ref{18}) referring to the Noether current gives
\begin{equation}\label{con}
\dot{B}^{3}f_{\mathcal{G}\mathrm{T}}=c_3,
\end{equation}
where $c_3$ is a constant of integration. It is to be mentioned that when we choose
$c_3$ equals to zero, this case satisfies the conservation
equation (\ref{18}). The corresponding Lagrangian becomes
\begin{align}\label{44_eqn}
\begin{split}
\mathcal{L}(B,R,\mathcal{G},\mathrm{\textit{T}},\dot{B},\dot{R},\dot{\mathcal{G}},
\dot{\mathrm{\textit{T}}})&=B^{m+2}[R-a_{0}\mathcal{G}^{2}-b_{0}\mathrm{\textit{T}}^{2}+2b_{0}\{\rho(B)\\&-
3p(B)\}\mathrm{\textit{T}}^{2}-p]-16a_{0}mB^{m-1}\dot{B}^{3}\dot{\mathcal{G}},
\end{split}
\end{align}
and the Euler-Lagrangian equations are calculated as
\begin{align}\label{42_eqn}
\begin{split}
&B^{m+2}\big(R-a_{0}\mathcal{G}^{2}-b_{0}T^{2}+2b_{0}(\rho(B)-3p(B))T-p\big)+32a_{0}m(m-1)\times\\&B^{m-2}\dot{B}^{3}\dot{\mathcal{G}}
+96a_{0}mB^{m-1}\ddot{B}\dot{\mathcal{G}}\dot{B}+48a_{0}mB^{m-1}\dot{B}^{2}\ddot{\mathcal{G}}=0,\\
\end{split}
\\&2a_{0}\mathcal{G}+16a_{0}m(m-1)B^{m-2}\dot{B}^{4}+48a_{0}m\dot{B}^{2}\ddot{B}B^{m-1}=0,\label{must}
\\&-2b_{0}B^{m+2}\big(T-(\rho(B)-3p(B))\big)=0.\label{musti}
\end{align}
Using Eq.(\ref{must}) and Eq.(\ref{musti}), it follows that
\begin{equation}\label{1214}
B=0,~~~~B=c_{4},
\end{equation}
and
\begin{equation}\label{1234}
B=c_{6}\big[(3m^{2}+3m+2)t-3c_{5}(1+m)\big]^\frac{3(1+m)}{2+3m+3m^{2}},
\end{equation}
where $c_{4}$, $c_{5}$ and $c_{6}$ are the constants of integration.
We discard the trivial solutions given in Eq.(\ref{1214}). For the non-trivial case (Eq.(\ref{1234})), the Ricci scalar $R$ takes the form as
\begin{eqnarray}
R&=&-\frac{6 (m+1) (7 m+5)}{\Big[(3 m (m+1)+2) t-3 c_5 (m+1)\Big]{}^2},
\label{Ricci}
\end{eqnarray}
while the GB term $\mathcal{G}$ is given as
\begin{eqnarray}
\mathcal{G}&=&-\frac{1296 m^3 (m+1)^3}{\Big[(3 m^2+3 m+2) t-3 c_{5} (m+1)\Big]^4}.
\end{eqnarray}
The average scale factor $a$ is calculated as
\begin{eqnarray}
a&=&\Big[c_{6} \Big\{(3 m^2+3 m+2) t-3 c_{5}(m+1)\Big\}^{\frac{3 (m+1)}{3 m^2+3 m+2}}\Big]^{\frac{1}{3} (2 m+1)}.
\end{eqnarray}
The 3D graphical behavior of the scale factor $a$ has been shown in Fig.1. It can be seen from both of the plots 1(a), and 1(b) that the scale factor is increasing with the passage of time, thus indicating the expansion in particular dimensions. This is because of the increase in the distance between cosmologically related objects as the time passes. This property is uniquely owned by the scale factor which is the function of cosmic time $t$. The limiting behavior of the scale factor will tend to infinity for the future era, provided this accelerating expansion of the universe continues.
\begin{figure}\center
\begin{tabular}{cccc}
& 1(a) & 1(b)\\ &
\epsfig{file=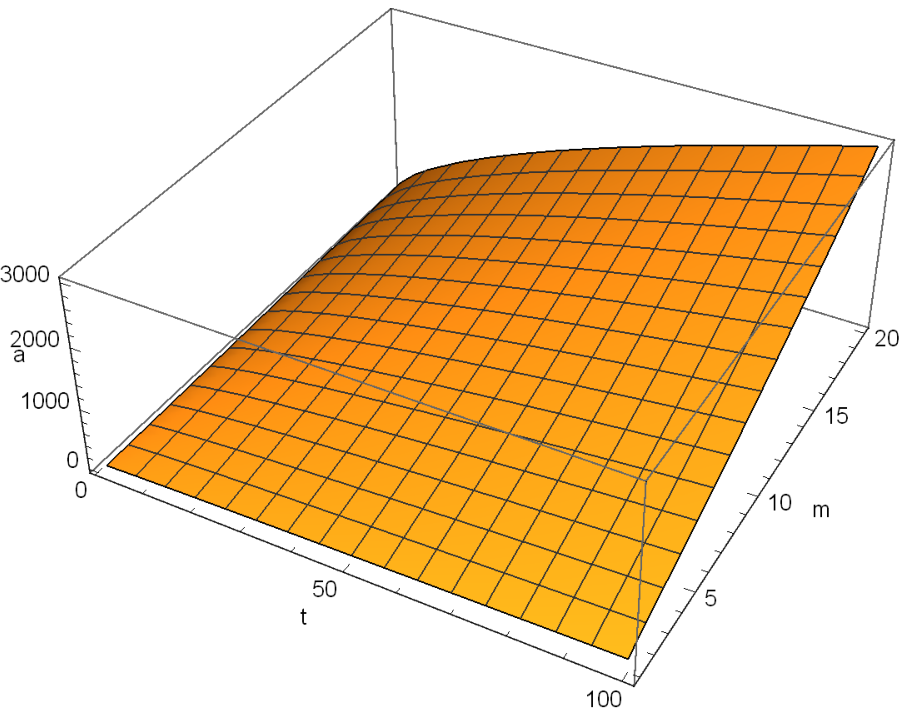,width=0.5\linewidth} &
\epsfig{file=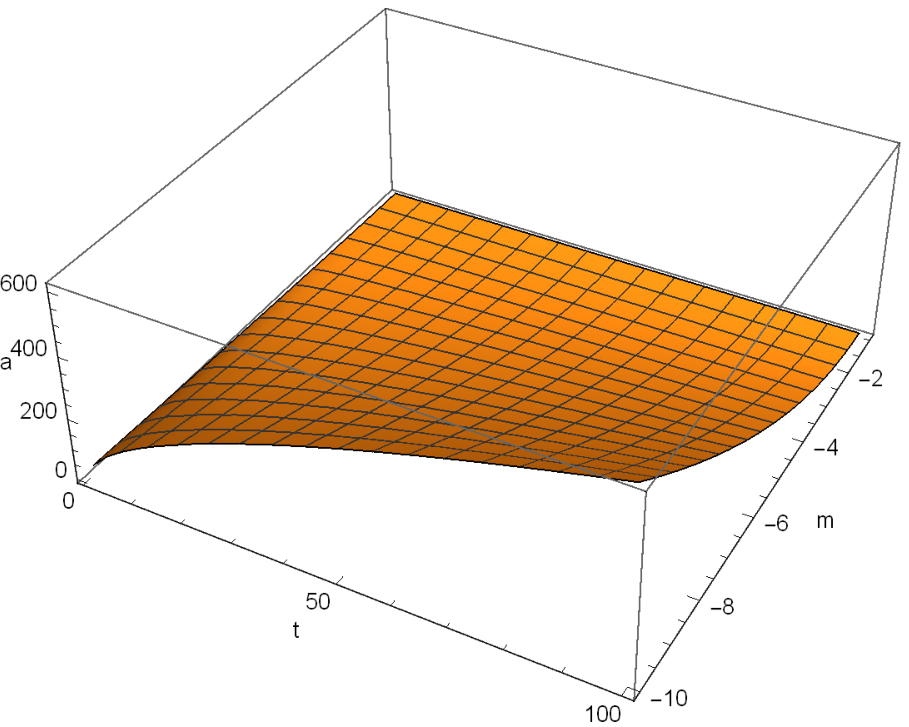,width=0.5\linewidth} \\
\end{tabular}
\caption{The plot of the scale factor $a(t)$ for positive values of $m$, Fig.1(a) and for the negative values of $m$, Fig.1(b).} \center
\end{figure}
Using Eqs.(\ref{eos}) and (\ref{42_eqn}-\ref{musti}), we get
\begin{equation}\label{45_eqn}
\rho^2(1-3\omega)^2-\omega\rho+l(t)=0,
\end{equation}
where
\begin{eqnarray}\nonumber
l(t)&=&-\frac{1679616m^6(1+m)^6}{\Big[(2+3m+3m^2)t-3(1+m)c_{5}\Big]^8}+\frac{1}{c_{6}}\Big\{4478976\times\\\nonumber
&&(m-1)m^4(1+m)^6(2+3m+3m^2)\Big[(2+3m+3m^2)t\\\nonumber
&&-3(1+m)c_{5}\Big]^{-8-\frac{3(1+m)}{2+3m+3m^2}}\Big\}-\frac{1}{c_{6}}\Big\{11197440m^4(1+m)^5\times\\\nonumber
&&(2+3m+3m^2)^2\Big[(2+3m+3m^2)t-3(1+m)c_{5}\Big]^{-8-\frac{3(1+m)}{2+3m+3m^2}}\Big\}\\\nonumber
&&+\frac{1}{c_{6}}\Big\{(4478976m^4(1+m)^5(2+3m+3m^2)^2(-1\\\nonumber
&&+\frac{3(1+m)}{2+3m+3m^2})\Big[(2+3m+3m^2)t-3(1+m)c_{5}\Big]^{-8-\frac{3(1+m)}{2+3m+3m^2}}\Big\}\\\nonumber
&&-\frac{6(1+m)(5+7m)}{\Big[(2+3m(1+m))t-3(1+m)c_{5}\Big]^2}.\\\nonumber
\end{eqnarray}
Now, one can see that Eqn.(\ref{45_eqn}) is quadratic in $\rho$ and its solution gives the following two roots
\begin{eqnarray}\label{46_eqn}
\rho &=&\frac{1}{2(1-3\omega)^2}\Big[\omega\pm\Big\{\omega^2+\frac{1}{\Big[(2+3m(1+m))t-3(1+m)c_5\Big]^8
c_{6}}\\\nonumber
&&\Big\{24(1-3\omega)^2\Big(746496m^4(1+m)^6(2+m-3m^3)\Big[(2+3m(1+m))t\\\nonumber
&&-3(1+m)c_{5}\Big]^{-\frac{3(1+m)}{2+3m(1+m)}}+746496m^4(1+m)^5(3m^2-1)\times\\\nonumber
&&(2+3m(1+m))\Big[(2+3m(1+m))t-3(1+m)c_{5}\Big]^{-\frac{3(1+m)}{(2+3m(1+m))}}\\\nonumber
&&+1866240m^4(1+m)^5(2+3m(1+m))^2\Big[(2+3m(1+m))t\\\nonumber
&&-3(1+m)c_{5}]\Big]^{-\frac{3(1+m)}{(2+3m(1+m))}}+279936m^6(1+m)^6c_{6}+(1+m)\times\\\nonumber
&&(5+7m)\Big[(2+3m(1+m))t-3(1+m)c_{5}\Big]^6c_{6}\Big)\Big\}\Big\}^{\frac{1}{2}}\Big].\\\nonumber
\end{eqnarray}
\begin{figure}\center
\begin{tabular}{cccc}
\epsfig{file=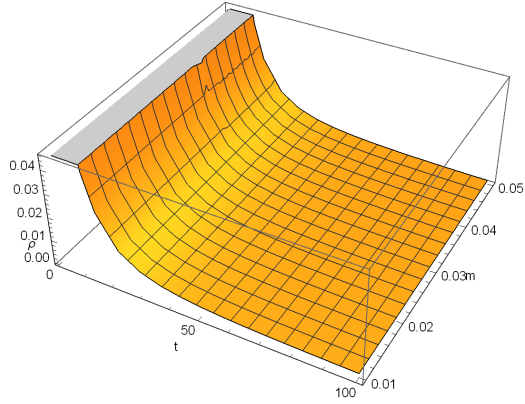,width=0.5\linewidth}
\end{tabular}
\caption{Behavior of the energy density $\rho(t)$ for the positive square root values when $\omega=-1$.} \center,
\end{figure}
The 3D graphical behavior of the energy density $\rho$ when plotted against the cosmic time $t$ is shown in the Fig.2. The plot clearly indicates the accelerating expansion of the universe when $\omega=-1$ with the positive value of the square root term. The EoS parameter $\omega$ describes the different epochs of the accelerating universe. For $-1<\omega\leq-1/3$, DE phase is divided by quintessence epoch whereas $\omega<-1$ and $\omega=-1$ correspond to phantom and cosmological constant $\Lambda$ eras respectively. Thus our solution corresponds to $\Lambda$CDM model which describes the parameterizations of the universe containing the cosmological constant $\Lambda$ associated with the DE. The graphical behavior of the first root of $\rho$ justifies the description of the $\Lambda$CDM model which exhibits the negative pressure $p=-\rho{c^2}$, and hence strengthens the reason of the accelerating expansion of the universe. Thus for this particular cosmological model, the solution metric takes the form
\begin{eqnarray}\nonumber
ds^{2}&=&d{t}^{2}-\Big[c_6 \Big\{(3 m^2+3 m+2) t-3 c_5 (m+1)\Big\}^{\frac{3 (m+1)}{3 m^2+3 m+2}}\Big]^{2 m}dx^{2}-\\
&&c_6^2 \Big[(3 m (m+1)+2) t-3 c_5 (m+1)\Big]^{\frac{6 (m+1)}{3 m (m+1)+2}}[d{y}^{2}+d{z}^{2}].
\end{eqnarray}

\section{Reconstruction of $\Lambda$CDM-Type Cosmology}

In this section, we propose a cosmological model $f(\mathcal{G},\textit{T})=\mathcal{G}^{k}\textit{T}^{1-k}$, where $k$ is any arbitrary real number. We develop some cosmological solutions for the case ${k}=2$. The Lagrangian
takes the form
\begin{equation}\label{41_eqn}
\mathcal{L}=B^{m+2}\Big[R-\frac{\{\rho(B)-3p(B)\}\mathcal{G}^{2}}{T^{2}}-p(B)\Big]-\frac{16B^{m-1}\dot{B}^{3}m}{T}\Big(\dot{\mathcal{G}}-\frac{\dot{T}}{T}\Big).
\end{equation}
Here, Euler-Lagrangian equations take the form
\begin{eqnarray}\nonumber
&&(m+2)B^{m+1}\Big[R-\frac{\{\rho(B)-3p(B)\}\mathcal{G}^{2}}{T^{2}}-p(B)\Big]-\frac{48mB^{m-1}\dot{B}^{2}\ddot{T}}{T^{2}}
\\\nonumber &&-B^{m+2}\Big[\frac{(\{\rho(B)-3p(B)\})_{,B}\mathcal{G}^{2}}{T^{2}}
+p_{,B}(B)\Big]+\frac{48m(m-1)B^{m-2}\dot{B}^{3}\dot{\mathcal{G}}}{T}
\\\nonumber &&+\frac{96m\dot{B}\ddot{B}B^{m-1}\dot{\mathcal{G}}}{T}+\frac{48mB^{m-1}\dot{B}^{2}\ddot{G}}{T}-\frac{48mB^{m-1}\dot{B}^{2}\dot{\mathcal{G}}\dot{T}}{T^{2}}
+\frac{96mB^{m-1}\dot{B}^{2}\dot{T}^{2}}{T^{3}}\\\nonumber&&-\frac{48m(m-1)B^{m-2}\dot{B}^{3}\dot{T}}{T^{2}}
\frac{96m\dot{B}\ddot{B}B^{m-1}\dot{T}}{T^{2}}-\frac{16m(m-1)B^(m-2)\dot{B}^{3}}{T}\Big[\dot{\mathcal{G}}-\frac{\dot{T}}{T}\Big]=0,\\&&\\\label{b}
&&\nonumber -2B^{m+2}(\rho(B)-3p(B))\mathcal{G}+16m(m-1)TB^{m-2}\dot{B}^{4}+48mTB^{m-1}\dot{B}^{2}\ddot{B}\\&&-16mB^{m-1}\dot{T}\dot{B}^{3}=0,\\\label{m}
&&\nonumber B^{m+2}(\rho(B)-3p(B))\mathcal{G}^{2}+8mTB^{m-1}\dot{B}^{3}\dot{\mathcal{G}}-8m(m-1)TB^{m-2}\dot{B}^{4}\\&&-24mB^{m-1}T\dot{B}^{2}\ddot{B}=0.\label{49_eqn}
\end{eqnarray}
By putting the corresponding values for $\mathcal{G}$ and
$\mathrm{\textit{T}}$ and using Eq.(\ref{49_eqn}), we get
\begin{equation}\label{41_eqn}
\begin{split}
&64m^{2}(m-1)(m+3)B^{m-6}\dot{B}^{8}+960B^{m-4}\dot{B}^{4}\ddot{B}^{2}+64m^{2}(10m-19)B^{m-5}\dot{B}^{6}\ddot{B}\\&+
192m^{2}\dot{B}^{5}\dot{B}^{m-4}\dddot{B}-8m(m-1)B^{m-2}\dot{B}^{4}-24mB^{m-1}\dot{B}^{2}\ddot{B}=0.
\end{split}
\end{equation}
This equation admits an exponential solution
\begin{equation}
B=e^{\varphi t},
\end{equation}
with the constraint equation
\begin{equation}\label{56}
8m^{2}(m^{2}+12m-4)\varphi^{8}-(m^{2}+2m)\varphi^{4}=0,
\end{equation}
where $\varphi$ is an arbitrary constant. Eq.(\ref{56}) yields the real solutions
\begin{equation}\label{phi1}
\varphi=0, ~~~\varphi=\pm\frac{(2+m)^{\frac{1}{4}}}{(-32m+96m^{2}+8m^{3})^{\frac{1}{4}}}.
\end{equation}
Now using the Eqs.$(\ref{b})$ and $(\ref{m})$, we get
\begin{equation}
\begin{split}
&B^{4}p_{,B}(B)+(m+2)B^{3}p(B)=\frac{1}{c_{7}}\big[48m(m-1)\dot{B}^{3}\dot{\mathcal{G}}
\\&+96m\dot{B}\ddot{B}B\dot{\mathcal{G}}+48mB\dot{B}^{2}\ddot{\mathcal{G}}-(m+2)B^{3}\mathcal{G}^{2}-16m(m-1)\dot{B}^{3}\dot{\mathcal{G}}\big].\label{d}
\end{split}
\end{equation}
It may be noted that Eq.(\ref{d}) is a non-homogeneous linear
differential equation in pressure $p(B)$ and its solution is obtained as
\begin{equation}
\begin{split}
p(B)=&c_{7}B^{-m-2}-\frac{192c_{8}m^{4}\dot{B}^{2}}{B^{8}}\Big[\frac{(136-152m+22m^{2}-9m^{3}+3m^{5})\dot{B}^{6}}{m-6}\\&+\frac{2(-182+154m-14m^{2}+3m^{3}+3m^{4})B\dot{B}^{4}\ddot{B}}{m-5}\\&+\frac{B^{2} \dot{B}^{2}\big((3 m^3+6 m^2-128 m+254)\ddot{B}^{2}+4 (16-7 m) B^{(3)} \dot{B}\big)}{m-4}\\&-\frac{6 B^{3}\big(6 \ddot{B}^{3}+B^{(4)}\dot{B}^{2}+8B^{(3)}\dot{B}\ddot{B}\big)}{m-3}\Big],
\end{split}
\end{equation}
where $c_{7}$ and $c_{8}$ are the constants of integration.
Now using the positive $\varphi$ from Eq.(\ref{phi1}), the expression for pressure explicitly becomes function of time
\begin{eqnarray}\\\nonumber
p(t)&=&c_{7}\exp \Big[\frac{(-m-2) \sqrt[4]{m+2} t}{\sqrt[4]{8 m^3+96 m^2-32 m}}\Big]-\frac{1}{[(m-6) (m-5) (m-4) (m-3)]}\times\\\nonumber
&&\Big[192 c_{8} \Big[\frac{\sqrt[4]{m+2}}{\sqrt[4]{8 m^3+96 m^2-32 m}}\Big]^8\times m^4 \Big[3 m^8-30 m^7+63 m^6+132 m^5\\\nonumber
&&-326 m^4-258 m^3-186 m^2+1094 m+228\Big]\Big].
\end{eqnarray}
\begin{figure}\center
\begin{tabular}{cccc}
& 3(a) & 3(b)\\ &
\epsfig{file=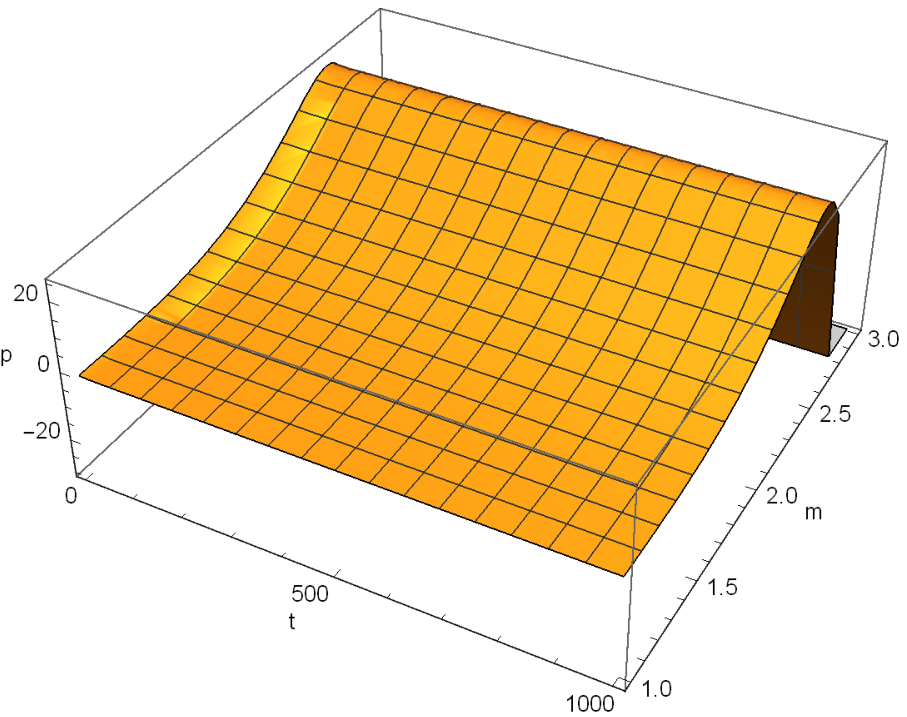,width=0.5\linewidth} &
\epsfig{file=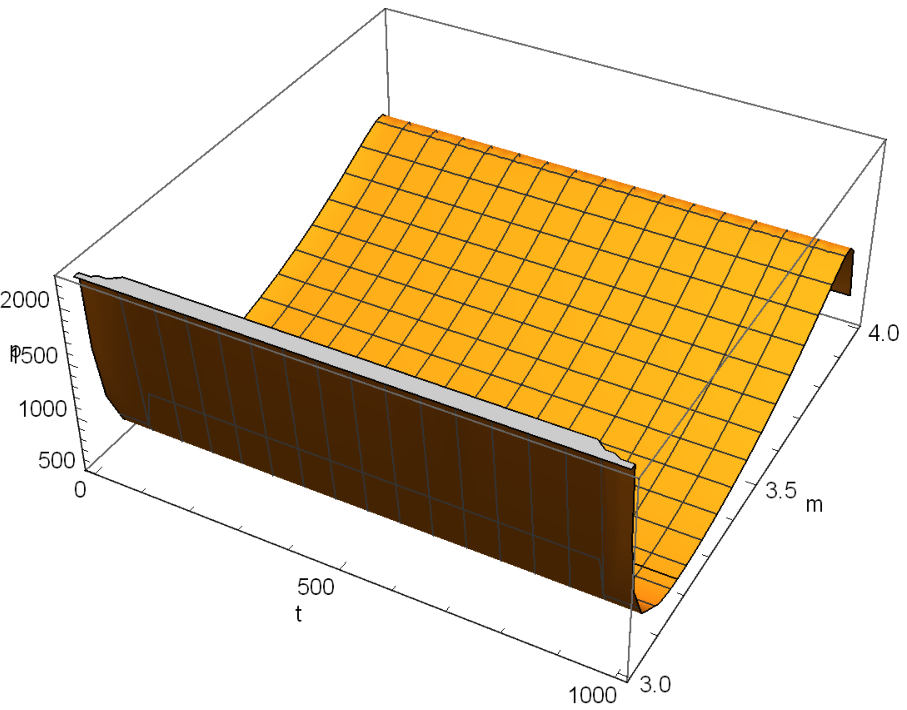,width=0.5\linewidth} \\
\end{tabular}
\caption{Behavior of the pressure $p(t)$, Fig.3(a) when $1<m<3$ and, Fig.3(b) when $3<m<4$, with $c_7=1$, $c_8=-1$.}\center
\end{figure}
\begin{figure}\center
\begin{tabular}{cccc}
& 4(a) & 4(b)\\ &
\epsfig{file=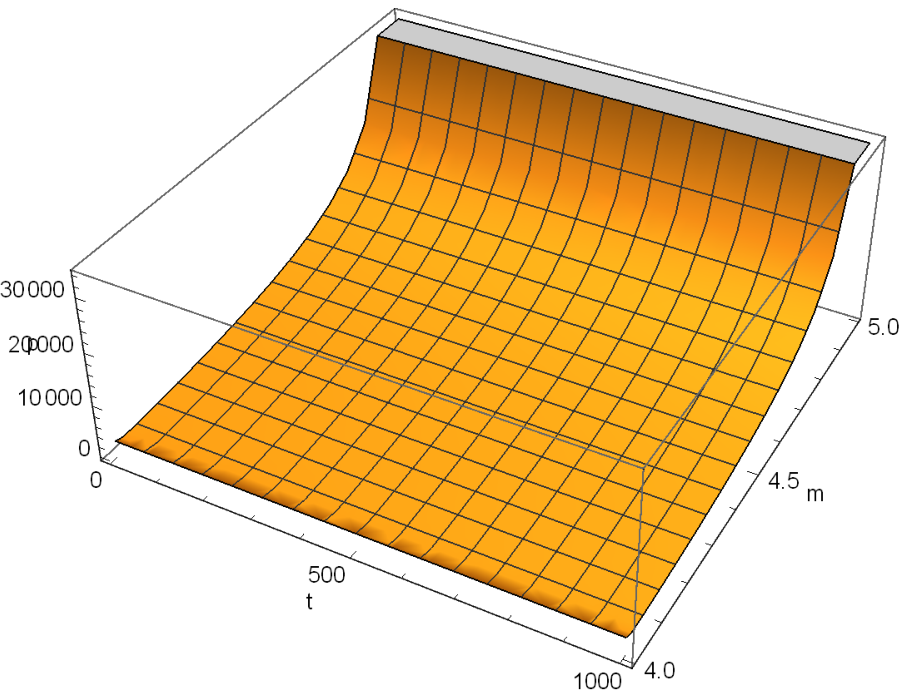,width=0.5\linewidth} &
\epsfig{file=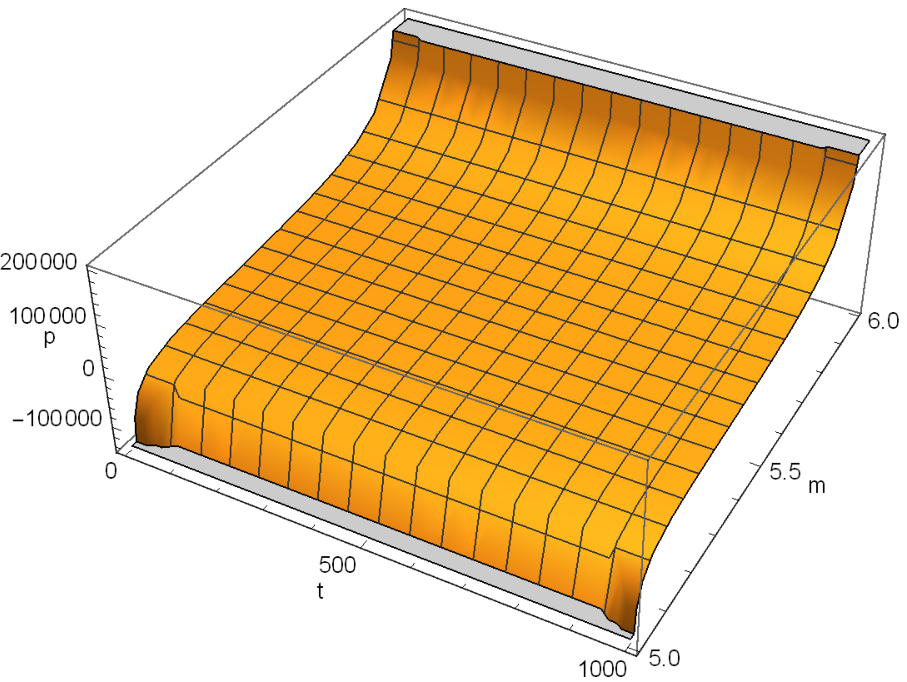,width=0.5\linewidth} \\
\end{tabular}
\caption{Behavior of the pressure $p(t)$, Fig.4(a) when $4<m<5$ and, Fig.4(b) when $5<m<6$, with $c_7=c_8=1$.}\center
\end{figure}
\begin{figure}\center
\begin{tabular}{cccc}
\epsfig{file=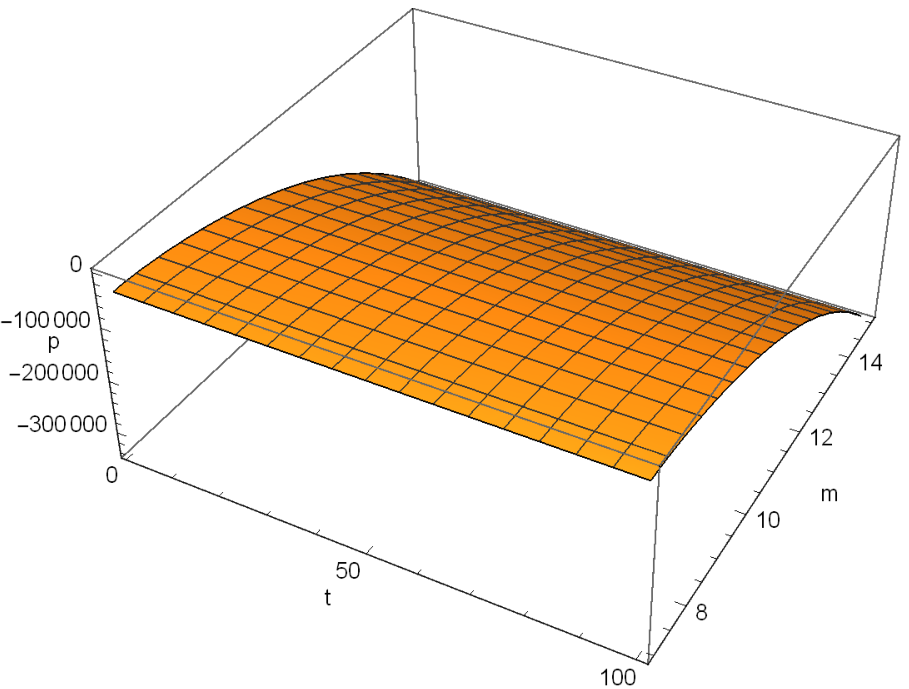,width=0.5\linewidth}
\end{tabular}
\caption{Behavior of the pressure $p(t)$, when $m>6$ with $c_7=c_8=1$.}\center
\end{figure}
The $3$D plotting of the pressure $p(t)$ against the time $t$ and anisotropy parameter $m$ shows the different behaviors of the pressure. The first four plots Fig.(3a-4b) have been avoided from the singularities caused due to $m={3,4,5,6}$, showing mostly the positive pressure but when $m>6$, the negative pressure is attained throughout as we go on increasing the time $t$ (as shown in Fig.(5)). Thus the solution for anisotropy parameter $m>6$ corresponds to $\Lambda$CDM model.
It is the interesting feature of modified GB gravity that specific $f(\mathcal{G},\textit{T})$ gravity models may be used to reconstruct $\Lambda$CDM cosmology without involving any cosmological constant.
Thus in this case the solution metric becomes
\begin{equation}\label{62a}
ds^{2}=d{t}^{2}-e^{2m\varphi t}dx^{2}-e^{2\varphi t}[d{y}^{2}+d{z}^{2}].
\end{equation}
The familiar deSitter space-time in GR is recovered when $m=1$. Here we have constructed a physical cosmological solution with a particular $f(\mathcal{G},\textit{T})$ gravity model. In the similar way, more solutions with some other cosmological models can be reconstructed.

\section{Concluding Comments}

This paper is devoted to study recently proposed $f(\mathcal{G},\mathrm{T})$ theory of gravity \cite{sharif.ayesha} with anisotropy background. For this purpose,
we consider LRS Bianchi type $I$ cosmological model in the presence of perfect fluid.  Since the field equations are highly non-linear and complicated, so we assume that ratio of shear and expansion scalars is constant, which gives $A=B^m$ \cite{ratio}. Noether symmetries not only aid
to investigate the hidden symmetries, but also their existence
provide suitable conditions so that we can choose physical
models of universe compatible with recent observations. Therefore, we analyze
Noether symmetries of the LRS Bianchi type $I$ universe in $f(\mathcal{G},\mathrm{T})$ theory of gravity.
It would be worthwhile to mention here that our results agree with \cite{shamir.ahmad} for a special case when $m=1$.
We have calculated the Lagrangian for LRS Bianchi type $I$ universe model in $f(\mathcal{G},\mathrm{T})$ theory. The
existence of Noether charges is extremely significant in the
literature, and the equation for conservation of charges plays an
important role to examine the Noether symmetries. The conservation
equation for Noether charges has been established.

The exact solutions of Noether equations have been discussed for two cases of
$f(\mathcal{G},\mathrm{T})$ gravity models. The first case when
$f_{\mathcal{G}\mathcal{G}}=0$ yields trivial symmetries while we
obtain non-trivial symmetries for the second case when
$f_{\mathcal{G}\mathrm{\textit{T}}}=0$ and
$f_{\mathcal{G}\mathcal{G}}\neq0$. Thus, the second case provides
$f(\mathcal{G},\textit{T})=a_{0}\mathcal{G}^{2}+b_{0}\textit{T}^{2}$
gravity model, where $a_{0}$ and $b_{0}$ are arbitrary constants.
Moreover, solutions in both cases satisfy the conservation equation
for Noether charges. We have also reconstructed some important cosmological solutions by
proposing $f(\mathcal{G},\textit{T})=\mathcal{G}^{k}\textit{T}^{1-k}$,
where $k$ is an arbitrary real number. This model yields the
familiar deSitter solution already available in GR. Moreover, the solutions for anisotropy parameter $m>6$ correspond to $\Lambda$CDM model.
Thus, the interesting feature of modified GB gravity is that specific $f(\mathcal{G},\textit{T})$ gravity models may be used to reconstruct $\Lambda$CDM cosmology without involving any cosmological constant.\\\\
\textbf{Acknowledgements}\\\\ The authors are thankful to
National University of Computer and Emerging Sciences (NUCES) for funding support.

\end{document}